\documentclass[preprint,aps,superscriptaddress]{revtex4}
\usepackage{graphicx,amsfonts}
\usepackage{amsmath}

\newcommand{\pa}{\partial}

\begin{document}

\title{On the consistency of the three-dimensional noncommutative supersymmetric Yang-Mills theory}

\author{A. F. Ferrari}
\author{H. O. Girotti}
\affiliation{Instituto de F\'{\i}sica, Universidade Federal do Rio Grande do
Sul, Caixa Postal 15051, 91501-970 - Porto Alegre, RS, Brazil}
\email{alysson, hgirotti, aribeiro@if.ufrgs.br}
\author{M. Gomes}
\author{A. Yu. Petrov}
\altaffiliation[]{Department of Theoretical Physics, Tomsk State Pedagogical University, Tomsk 634041, Russia (e-mail: petrov@tspu.edu.ru)}
\affiliation{Instituto de F\'{\i}sica, Universidade de S\~{a}o Paulo, Caixa
Postal 66318, 05315-970, S\~{a}o Paulo - SP, Brazil}
\email{mgomes, petrov, ajsilva@fma.if.usp.br}
\author{A. A. Ribeiro}
\affiliation{Instituto de F\'{\i}sica, Universidade Federal do Rio Grande do
Sul, Caixa Postal 15051, 91501-970 - Porto Alegre, RS, Brazil}
\author{A. J. da Silva}
\affiliation{Instituto de F\'{\i}sica, Universidade de S\~{a}o Paulo, Caixa
Postal 66318, 05315-970, S\~{a}o Paulo - SP, Brazil}

\begin{abstract}
We study the one-loop quantum corrections to the $U(N)$ noncommutative
supersymmetric Yang-Mills theory in three spacetime dimensions (NCSYM$_3$). We show that the cancellation of the dangerous UV/IR infrared divergences only takes place in the fundamental representation of the gauge group. Furthermore, in the one-loop approximation, the would be subleading UV and UV/IR infrared divergences are shown to vanish. 
\end{abstract}

\maketitle

Noncommutative supersymmetric models have a prominent place among the physically interesting field theories. Because supersymmetry favours the cancellation of dangerous divergences, they are the best candidates in a program to define consistent noncommutative field theories \cite{Mat, bichl, Zanon, Girotti1}. As part of a sequence of investigations devoted to this question \cite{ours, 3qed, cpn, ncsym4d}, in this paper we use the covariant superfield approach to study the noncommutative supersymmetric Yang-Mills model in $2+1$ spacetime dimensions. Based upon our previous experience with noncommutative supersymmetric QED$_3$ \cite{3qed}, we expect the absence of all one-loop divergences. More precisely, we shall show that the cancellation of the harmful (linear) UV/IR infrared divergences is achieved for the $U(N)$ group but only in the fundamental representation. We also verify the absence of the UV and UV/IR infrared logarithmic divergences for the entire theory in the one-loop approximation.

A general analysis of the divergence structure unveils that we must only care about the two-point vertex function of the spinor gauge superpotential. Indeed, although by power counting the three- and four-point vertex functions of this field are logarithmically divergent, they are in fact finite due to symmetric integration.
 
The action of the NCSYM$_3$ theory is \cite{SGRS}

\begin{eqnarray}
\label{asym}
S\,=\,\frac{1}{2g^2}{\rm Tr}\int d^5 z \, W^\alpha *W_\alpha\,, \label{2n}
\end{eqnarray}

\noindent
where

\begin{eqnarray}
\label{sstr}
W_\beta \,=\,\frac{1}{2}D^\alpha D_\beta \Gamma_\alpha -
\frac{i}{2}[\Gamma^\alpha ,D_\alpha \Gamma_\beta ]-
\frac{1}{6}
[\Gamma^\alpha ,\{\Gamma_\alpha ,\Gamma_\beta \}]\,,
\end{eqnarray}

\noindent
is the superfield strength constructed from the spinor
superpotential $\Gamma_\alpha $ which is a Lie-algebra valued superfield,
$\Gamma_{\alpha}(z)\,=\,\Gamma_{\alpha}^a (z) T^a,\, a = 1, \ldots, N^2$. 
Hereafter, it is assumed that all commutators and anticommutators are Moyal ones. 
The above action is invariant under the $U(N)$ infinitesimal gauge transformation

\begin{eqnarray}
\label{gt}
\delta \Gamma_{\alpha}\,=\,D_{\alpha}K-i[\Gamma_{\alpha},K]\,,
\end{eqnarray}

\noindent
where $K(z)=K^a (z) T^a$ is a supergauge parameter. 

A generic covariant gauge ($\xi$) is fixed by adding to Eq.~(\ref{asym}) the term

\begin{eqnarray}
S_{GF}\,=\,-\frac{1}{4\xi g^2}{\rm Tr}\int d^5 z (D^\alpha \Gamma_\alpha )
D^2(D^\beta \Gamma_\beta )\,.
\end{eqnarray}

\noindent
One is also to include the Faddeev-Popov action 

\begin{eqnarray}
\label{sfp}
S_{FP}\,=\,\frac{1}{2g^2}{\rm Tr}\int d^5 z (c'D^\alpha D_\alpha c + ic'*
D^\alpha [\Gamma_\alpha ,c])\,,
\end{eqnarray}

\noindent
where the ghost fields $c$ and $c'$ are also Lie-algebra valued superfields.

Altogether, the resulting quadratic part of the action reads

\begin{eqnarray}
\label{s2a}
S_2\,=\,\frac{1}{2g^2}{\rm Tr}\int d^5 z
\Big[\frac{1}{2}(1+\frac{1}\xi)\Gamma^\alpha
\Box \Gamma_\alpha
-\frac{1}{2}(1-\frac{1}\xi )\Gamma^\alpha i\pa_{\alpha\beta}
D^2\Gamma^\beta \,+\, c'D^\alpha D_\alpha c \Big]\,,
\end{eqnarray}

\noindent
leading to the free propagators

\begin{eqnarray}
<\Gamma^{\alpha a} (z_1)\Gamma^{\beta b} (z_2)>\,=\,
{ig^2}\delta^{ab}\left[
C^{\alpha\beta}\frac{1}{\Box}(\xi+1)-\frac{1}{\Box^2}
(\xi-1)i\pa^{\alpha\beta}D^2\right]\delta^5(z_1-z_2)
\end{eqnarray}

\noindent
and

\begin{eqnarray}
\label{pr2}
<c^{\prime a}(z_1)c^b(z_2)>\,=\,-2ig^2\delta^{ab}\frac{D^2}{\Box}
\delta^5(z_1-z_2)\,,
\end{eqnarray}

\noindent 
where $C^{\alpha\beta}=-C_{\alpha\beta}$ is the second-rank
antisymmetric tensor with normalization $C^{12}=i$. Furthermore, we take

\begin{eqnarray}
{\rm Tr}(T^aT^b)\,=\,\frac{1}{2}\delta^{ab},
\end{eqnarray}

\noindent
corresponding to the fundamental representation of the group generators. As known, at the classical level the use of this representation is mandatory  to guarantee the closure of the gauge algebra \cite{Chaichian,Terashima,Matsu}.

The interacting part of the action is

\begin{eqnarray}
\label{sint}
S_{\mbox{int}}&=&\frac{1}{g^2}{\rm Tr}\int d^5 z\Big\{-\frac{i}{4}D^{\gamma}D^\alpha
\Gamma_{\gamma}*
[\Gamma^\beta ,D_\beta \Gamma_\alpha ]-\frac{1}{12}D^{\gamma}D^\alpha \Gamma_{\gamma}*
[\Gamma^\beta ,\{\Gamma_\beta ,\Gamma_\alpha \}]-\nonumber\\&-&
\frac{1}{8}[\Gamma^{\gamma},D_{\gamma}\Gamma^\alpha ]*[\Gamma^\beta ,D_\beta \Gamma_\alpha
]+\frac{i}{12}
[\Gamma^{\gamma},D_{\gamma}\Gamma^\alpha ]*[\Gamma^\beta ,\{\Gamma_\beta ,\Gamma_\alpha \}]+
\nonumber\\&+&\frac{1}{72}
[\Gamma^{\gamma},\{\Gamma_{\gamma},\Gamma^\alpha \}]*[\Gamma^\beta ,\{\Gamma_\beta ,\Gamma_\alpha \}]\,+\,\frac{i}{2} c'*D^\alpha [\Gamma_\alpha ,c]
\Big\}\,.
\end{eqnarray}

One then may convince oneself that the superficial degree of divergence for this model is 

\begin{eqnarray}
\label{o}
\omega\,=\,2-\frac{1}{2}V_c-2V_0-\frac{3}{2}V_1-V_2-\frac{1}{2}V_3-\frac{1}{2}N_D\,,
\end{eqnarray}

\noindent
where $V_i,\, i = 0, \ldots, 3$, is the number of pure gauge vertices involving $i$ spinor supercovariant derivatives, $V_c$ is the number of ghost vertices and $N_D$ is the number of spinor derivatives moved to the external lines as consequence of the D-algebra transformations. By invoking the topological relation linking the numbers of loops, vertices and internal lines, one can convince oneself that all diagrams beyond the two-loop order are superficially finite. Hence, the theory under analysis is super-renormalizable. As we shall see, the same remains true after the introduction of matter superfields.

\begin{figure}[ht]
\includegraphics{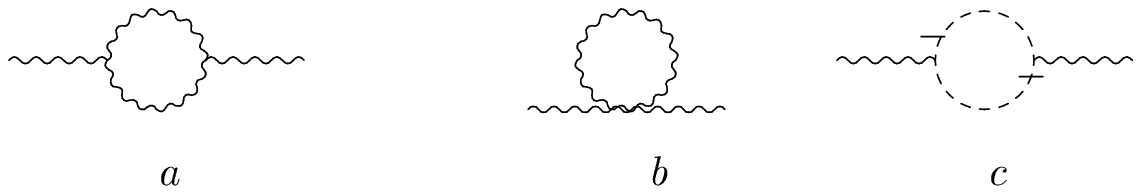}
\caption{\label{3d} Diagrams contributing to the two-point function of the
gauge superfield: gauge sector.}
\end{figure}

From Eq.~(\ref{o}) follows that the only linearly divergent graphs are those depicted in 
Fig.~\ref{3d}, where wavy and dashed lines represent gauge and ghost free propagators, respectively. They yield planar and nonplanar contributions. The linear UV divergences in the planar sectors are washed out by dimensional regularization, while the UV logarithmic ones vanish by symmetric integration. As for the nonplanar contributions, they are found to read

\begin{subequations}
\label{contrib1}
\begin{eqnarray}
I_a&=&\xi\int \frac{d^3p}{(2\pi)^3}d^2 \theta_1
\int\frac{d^3k}{(2\pi)^3}\frac{\cos(2k\wedge p)}{k^2}
\Gamma^{b\beta} (-p,\theta_1)\Gamma^{b^{\prime}}_\beta (p,\theta_1)
A^{acb}A^{acb^{\prime}}
\,+\,\cdots\,,\label{mlett:acontrib1}\\
I_b&=&-\frac{1+\xi}{2}\int \frac{d^3p}{(2\pi)^3}d^2 \theta_1
\int\frac{d^3k}{(2\pi)^3}\frac{\cos(2k\wedge p)}{k^2}
\Gamma^{b\beta} (-p,\theta_1)\Gamma^{b^{\prime}}_\beta (p,\theta_1)
A^{abab^{\prime}}
+\cdots ,\label{mlett:bcontrib1}\\
I_c&=&\int \frac{d^3p}{(2\pi)^3}d^2 \theta_1
\int\frac{d^3k}{(2\pi)^3}\frac{\cos(2k\wedge p)}{k^2}
\Gamma^{b\beta} (-p,\theta_1)\Gamma^{b^{\prime}}_\beta (p,\theta_1)
A^{acb}A^{acb^{\prime}}\,+\,\cdots\,,\label{mlett:ccontrib1}
\end{eqnarray}
\end{subequations}

\noindent
where 

\begin{equation}
A^{a_1 a_2 \cdots a_n}={\rm Tr}(T^{a_1}T^{a_2}\cdots T^{a_n})\,,
\end{equation}

\noindent
$k \wedge p \, \equiv \, \frac{1}{2} k_i \Theta^{ij} p_j$ and $\Theta^{ij}$ is the antisymmetric matrix characterizing the noncommutativity of the underlying spacetime \cite{footnote1}. The presence of harmful (linear) UV/IR infrared singularities in Eq.~(\ref{contrib1}) should be noticed. They are accompanied by subleading (harmless) UV/IR infrared singularities as well as by finite terms, both of them indicated by dots. It is worth mentioning that these subleading UV/IR infrared singularities arise from integrals of the kind ($\tilde{p}^i \,\equiv\,\Theta^{ij}p_j$)

\begin{eqnarray}
\label{typint}
\int\frac{d^3k}{(2\pi)^3}\frac{k^j e^{2ik\wedge p}}{k^4}=-\frac{i}{4\pi}
\frac{\tilde{p}^j}{\sqrt{\tilde{p}^2}}\,,
\end{eqnarray}

\noindent
which remains finite although depending on the direction as ${\tilde p} \rightarrow 0$. 

From Eq.~(\ref{contrib1}) we find that the linear UV/IR infrared divergences are cancelled if and only if 

\begin{eqnarray}
\label{cond}
{\rm Tr}(T^bT^aT^bT^c)\,=\,2\,{\rm Tr}(T^bT^dT^a){\rm Tr}(T^bT^dT^c)\,,
\end{eqnarray}

\noindent
which is indeed verified in the fundamental representation of the gauge group. To see that this is the case, one may use the completeness relation for the $U(N)$ generators in the fundamental representation

\begin{equation}
\label{III-81}
(T^a)_{ij}(T^a)_{kl}\,=\,\frac{1}{2}\,\delta_{il}\delta_{jk}\,.
\end{equation}

\noindent
When compared with the situation encountered in noncommutative supersymmetric QED$_3$ \cite{3qed}, one sees that Eq.~(\ref{cond}) constitutes a new requirement arising from  the non-Abelian character of the gauge group. Surprisingly, the same requirement secures the absence of quadratic UV/IR infrared divergences in the non-Abelian four-dimensional case \cite{ncsym4d}, in spite of the strong differences in the structures of these theories.

\begin{figure}[ht]
\includegraphics{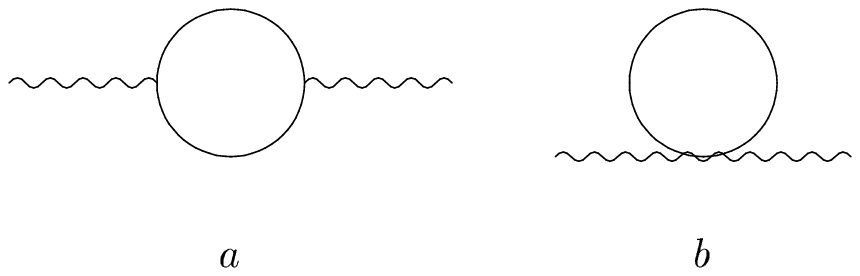}
\caption{\label{3dm} Diagrams contributing to the two-point function of
the gauge superfield: matter sector.}
\end{figure}

We shall next introduce matter minimally coupled to the gauge superfield. As consequence, the pure gauge action becomes augmented by the term

\begin{eqnarray}
\label{sm}
S_M&=&{\rm Tr}
\int d^5 z \left[\frac{1}{2}(D^{\alpha} \bar{\phi}+i[\bar{\phi},\Gamma^{\alpha}])*
(D_{\alpha} \phi-i[\Gamma_{\alpha} ,\phi])+m\bar{\phi} \phi
\right]\,,
\end{eqnarray}

\noindent
where $\phi (z) = \phi^a (z) T^a$. The superficial degree of divergence is now given by

\begin{eqnarray}
\label{o1}
\omega=2-\frac{1}{2}V_c-2V_0-\frac{3}{2}V_1-V_2-\frac{1}{2}V_3-
\frac{1}{2}E_{\phi}-\frac{1}{2}V^D_{\phi}-V^0_{\phi}-\frac{1}{2}N_D\,,
\end{eqnarray}

\noindent
where $E_{\phi}$ is the number of scalar legs, and $V^D_{\phi}$ ($V^0_{\phi}$)
is the number of vertices involving scalar superfields with one (none)
spinor supercovariant derivatives. By power counting, the linearly divergent supergraphs involving matter vertices are those depicted in Fig.~\ref{3dm}, where the continuous line represents the free matter field propagator

\begin{equation}
\label{m1}
<\bar{\phi}^a(z_1)\phi^b(z_2)>\,=\,
-\,2\,i \,\delta^{ab}\,\frac{D^2+m}{\Box-m^2}\delta^5(z_1-z_2)\,.
\end{equation}

\noindent
After D-algebra transformations, the graph in Fig.~2a gives the contribution

\begin{eqnarray}
\label{s1}
I_{2a}&=&-4 \int \frac{d^3p}{(2\pi)^3} d^2\theta \int \frac{d^3k}{(2\pi)^3}
\frac{\cos(2k\wedge p)}{(k^2+m^2)\left[(k+p)^2+m^2\right]}
\nonumber\\&\times&\Big[-
(k^2+m^2)C_{\alpha\beta} \Gamma^{\alpha b}(-p,\theta)
\Gamma^{\beta b^{\prime}}(p,\theta)
+(k_{\alpha\beta}-mC_{\alpha\beta})
\left[D^2\Gamma^{\alpha}(-p,\theta)\right] \Gamma^{\beta}(p,\theta)
\nonumber\\&+&{\frac{1}{2}} D^{\gamma}D^{\alpha}\Gamma_{\alpha}
(k_{\gamma\beta}-mC_{\gamma\beta})\Gamma^{\beta}(p,\theta)
\Big]{\rm Tr}(T^aT^cT^b){\rm Tr}(T^aT^cT^{b\prime})+\ldots\,,
\end{eqnarray}

\noindent
whereas the graph in Fig.~2b yields

\begin{eqnarray}
I_{2b}&=&-2 \int \frac{d^3p}{(2\pi)^3} d^2\theta
\int \frac{d^3k}{(2\pi)^3}\frac{\cos(2k\wedge p)}{(k+p)^2+m^2}
C_{\alpha\beta} 
\Gamma^{\alpha b}(-p,\theta)\Gamma^{\beta b^{\prime}}(p,\theta)
\times\nonumber\\&\times&
{\rm Tr}(T^aT^bT^aT^{b^{\prime}})+\ldots\,.
\end{eqnarray}

We see again that the linear divergences are cancelled if and only if condition (\ref{cond})
is satisfied. As before, logarithmic divergences vanish due to symmetric momentum integration.
Therefore, the completeness of the fundamental representation of the gauge group, expressed in Eq.~(\ref{III-81}), secures the one-loop finiteness of NCSYM$_3$ coupled to matter. Also, Eq.~(\ref{o1}) implies in the absence of divergent corrections beyond two-loop order.

If, instead of adding Lie algebra valued matter superfields, we take matter to be a $N$ component column vector, namely, in the fundamental representation, the corresponding part of the action reads

\begin{eqnarray}
\label{smf}
S_M&=& \int d^5 z \left[{\frac{1}{2}}(D^{\alpha} \bar{\phi}+i\bar{\phi}\Gamma^{\alpha})*
(D_{\alpha} \phi-i\Gamma_{\alpha} \phi)+m\bar{\phi} \phi
\right]\,.
\end{eqnarray}

\noindent
In this case, the supergraphs in Fig.~2 would be totally planar and hence finite in the
framework of the dimensional regularization. 

We conclude that NCSYM$_3$ is a consistent theory (without nonintegrable UV/IR infrared singularities) and a sound candidate for enlarging the class of finite noncommutative field theories proposed in \cite{JJ}.

\vspace{1cm}

{\bf Acknowledgments.}

This work was partially supported by Funda\c{c}\~{a}o de Amparo 
\`{a} Pesquisa do Estado de S\~{a}o Paulo (FAPESP) and Conselho 
Nacional de Desenvolvimento Cient\'{\i}fico e Tecnol\'{o}gico (CNPq). 
H. O. G. also acknowledges support from PRONEX
under contract CNPq 66.2002/1998-99. A. Yu. P. has been supported by
FAPESP, project No. 00/12671-7.

\newpage

\begin{thebibliography}{99}

\bibitem{Mat} A. Matusis, L. Susskind, N. Toumbas, JHEP {\bf 12} (2000) 002.

\bibitem {bichl} A. A. Bichl, M. Ertl, A. Gerhold, J. M. Grimstrup, H. Grosse, L. Popp, V. Putz, M. Schweda, R. Wulkenhaar, ``Non-commutative U(1) Super-Yang-Mills Theory: Perturbative
Self-Energy Corrections'', hep-th/0203141.

\bibitem{Zanon} D. Zanon, Phys. Lett. {\bf B502} (2001) 265; Phys. Lett. {\bf B504} (2001) 101; M. Pernici, A. Santambrogio, D. Zanon, Phys. Lett. {\bf B504} (2001) 131; A. Santambrogio, D. Zanon, JHEP {\bf 01} (2001) 024.

\bibitem{Girotti1} H. O. Girotti, M. Gomes, V. O. Rivelles, A. J. da Silva, Nucl. Phys. {\bf B587} (2000) 299.

\bibitem{ours} A. F. Ferrari, H. O. Girotti, M. Gomes, A. Yu. Petrov, A. A. Ribeiro, V. O. Rivelles, A. J. da Silva, Phys. Rev. {\bf D69} (2004) 025008.

\bibitem{3qed}  A. F. Ferrari, H. O. Girotti, M. Gomes, A. Yu. Petrov, A. A. Ribeiro, A. J. da Silva, Phys. Lett. {\bf B577} (2003) 83.

\bibitem{cpn} E. A. Asano, H. O. Girotti, M. Gomes, A. Yu. Petrov, A. G. Rodrigues, A. J. da Silva, Phys.Rev. {\bf D69} (2004) 105012.

\bibitem{ncsym4d} A. F. Ferrari, H. O. Girotti, M. Gomes, A. Yu. Petrov, A. A. Ribeiro, V. O. Rivelles, A. J. da Silva, ``Towards a consistent noncommutative supersymmetric Yang-Mills theory: superfield covariant analysis'', hep-th/0407040.

\bibitem{SGRS} S. J. Gates, M. T. Grisaru, M. Rocek, W. Siegel, {\it Superspace or One Thousand and One Lessons in Supersymmetry} (Benjamin/Cummings, 1983).

\bibitem{Chaichian} M. Chaichian, P. Presnajder, M. M. Sheikh-Jabbari, and A. Tureanu, Phys. Lett. {\bf B526} (2002) 132.

\bibitem{Terashima} S. Terashima, Phys. Lett. {\bf B482} (2000) 276.

\bibitem{Matsu} K. Matsubara, Phys. Lett. {\bf B482} (2000) 417.

\bibitem{footnote1} We shall restrict to the case $\Theta^{0i}=0$ to avoid unitarity problems.

\bibitem{JJ} I. Jack, D. R. T. Jones, Phys. Lett. {\bf B514} (2001) 401; New. J. Phys. {\bf 3} (2001) 19.

\end{thebibliography}
\end{document}